\documentclass[12pt]{article}
\def\beq{\begin{equation}}
\def\eeq{\end{equation}}
\def\ap#1#2#3 {Ann. Phys. (NY) {\bf#1} (#2) #3}
\def\err#1#2#3 {{\it Erratum} {\bf#1} (#2) #3}
\def\ib#1#2#3 {{\it ibid.} {\bf#1} (#2) #3}
\def\ijmp#1#2#3 {Int. J. Mod. Phys. {\bf#1} (#2) #3}
\def\jetp#1#2#3 {JETP Lett. {\bf#1} (#2) #3}
\def\mpl#1#2#3 {Mod. Phys. Lett. {\bf#1} (#2) #3}
\def\np#1#2#3 {Nucl. Phys. {\bf#1} (#2) #3}
\def\pl#1#2#3 {Phys. Lett. {\bf#1} (#2) #3}
\def\prep#1#2#3 {Phys. Rep. {\bf#1} (#2) #3}
\def\prev#1#2#3 {Phys. Rev. {\bf#1} (#2) #3}
\def\prl#1#2#3 {Phys. Rev. Lett. {\bf#1} (#2) #3}
\def\sjnp#1#2#3 {Sov. J. Nucl. Phys. {\bf#1} (#2) #3}
\def\spj#1#2#3 {Sov. Phys. JETP {\bf#1} (#2) #3}
\def\spu#1#2#3 {Sov. Phys. Usp. {\bf#1} (#2) #3}
\def\zp#1#2#3 {Zeit. Phys. {\bf#1} (#2) #3}

\begin{document}
\begin{titlepage}
\begin{center}
{\Large \bf Theoretical Physics Institute \\
University of Minnesota \\}  \end{center}
\vspace{0.2in}
\begin{flushright}
TPI-MINN-01/10-T \\
UMN-TH-1942-01 \\
February 2001 \\
\end{flushright}
\vspace{0.3in}
\begin{center}
{\Large \bf  Once again on electromagnetic properties of a domain
wall interacting with charged fermions
\\}
\vspace{0.2in}
{\bf M.B. Voloshin  \\ }
Theoretical Physics Institute, University of Minnesota, Minneapolis,
MN
55455 \\ and \\
Institute of Theoretical and Experimental Physics, Moscow, 117259
\\[0.2in]
\end{center}

\begin{abstract}
The response to a magnetic flux is considered of the vacuum state of
charged Dirac fermions interacting with a domain wall made of a neutral
spinless field  in (3+1) dimensions with the fermion mass having a phase
variation across the wall.  It is pointed out that due to simple $C$
parity arguments the spontaneous magnetization for this system is
necessarily zero, thus invalidating some claims to the contrary in the
literature. The cancellation of the spontaneous magnetization is
explicitly demonstrated in a particular class of models. The same
calculation produces a general formula for the electric charge density
induced by the magnetic flux -- an effect previously discussed in the
literature for axionic domain walls. The distribution of the induced
charge is calculated in specific models.
\end{abstract}

\end{titlepage}

\section{Introduction}

The possibility that domain walls that could have existed in the early
universe could also be related to generation of a primordial magnetic
field correlated at large distances \cite{gr} has been recently
discussed in the literature [2-5]. The discussed models are based on the
idea that fermions, coupled to the field forming the wall, develop a
spontaneous magnetization perpendicular to the wall. Although it is a
simple exercise in general physics to show that a uniform magnetization
of an infinite domain wall does not produce magnetic field \cite{mv},
the phenomenon of magnetization of the wall is interesting on its own.

The claims to a non-zero magnetic moment of the ground state of a
fermion field coupled to the wall in (3+1) dimensions are inferred from
the behavior  in a (2+1) dimensional QED of Dirac fermions with a
definite sign of the mass term $m$. Namely, in certain calculations
\cite{iwazaki,cea,hosotani,ct2} of the energy of the ground state of the
fermion field in an external magnetic field $B$, it is claimed that the
total energy contains a linear in magnetic field term proportional to $m
\, B$, which corresponds to a spontaneous magnetization proportional to
the mass parameter $m$ (including the sign). For a Dirac fermion field
coupled to a domain wall in (3+1) dimensions the quantization of the
motion perpendicular to the wall splits the fermion system into an
infinite set of modes, each corresponding to a (2+1) dimensional QED
with its own parameter $m$. If the phase of the fermion mass varies
across the wall, the set of positive values of $m$ differs from that of
the negative values of $m$. Therefore in this picture it might  at least
be not obvious that the overall magnetization cancels after summation
over modes corresponding to positive and negative values of $m$.

It is nevertheless quite easy to argue that the cancellation necessarily
takes place and the magnetization of the fermion field ground state at
the wall in (3+1) dimensions is strictly zero. Indeed, the Lagrangian
density for the fermions with the phase of the mass $\mu$ depending on
the coordinate $z$ perpendicular to the wall can be generally written as
\beq
L={\overline \psi} \,\left [ i \, \left( \partial_\alpha - i \, A_\alpha
\right ) \, \gamma^\alpha - \mu_1 (z) - i \, \mu_2 (z) \, \gamma^5
\right ] \, \psi
\label{lgen}
\eeq
with $\mu_1$ and $\mu_2$ being respectively the real and imaginary parts
of $\mu(z)$, and $A$ standing for the electromagnetic field potential
and absorbing the charge $e$ in the normalization of the field. The
variation of $\mu_2$ breaks the $P$ and $CP$ parities. However the $C$
parity is manifestly conserved with both $\mu_1$ and $\mu_2$ being
$C$-even. Therefore after the fermions are integrated out the energy of
the system as a function of $A$ in the $C$-even background of $\mu(z)$
cannot contain odd powers of the $C$-odd field $A$. In particular the
energy cannot be linear in the magnetic field $B$. The same argument
holds also for the situation where the fermions are assumed to possess
anomalous magnetic and/or electric dipole moment since both such
interactions also conserve the $C$ parity.

This simple general $C$ parity argument is clearly sufficient for
excluding the possibility of spontaneous magnetization of the fermion
vacuum in a domain wall background. However due to the existence in the
literature of claims to the contrary \cite{ct,ct2} it is quite
instructive to demonstrate explicitly the vanishing of the linear in the
magnetic field $B_z=B$ term in the energy (as well as of all the odd
terms) at least in a specific model. Moreover the $C$ parity argument
certainly allows dependence of the energy on {\it even} powers of the
electromagnetic field, which gives rise to the most interesting
phenomenon of appearance of electric charge density once a magnetic flux
is applied across the wall. This phenomenon is related to the well known
coupling of a pseudoscalar field to the electromagnetic invariant ${\bf
E}\cdot{\bf B }$. For slowly varying with $z$ mass term, one can
approximate $\mu(z)$ around a given point $z_0$ as $\mu(z) = \mu(z_0) +
\delta \mu(z)$, and treating the varying part  $\delta \mu(z)$ as small
perturbation, find the term proportional to ${\bf E}\cdot{\bf B }$ in
the
energy density $w$ from the well known triangle graph as
\beq
\delta w =  {1 \over 4 \pi^2} \,{ \mu_2 \,\delta \mu_1 - \mu_1 \, \delta
\mu_2 \over {\mu_1^2 + \mu_2^2} } \, {\bf E}\cdot{\bf B}~.
\label{triag}
\eeq
The charge density is then found from variation with respect to the
potential $A_0$. For the magnetic field in the $z$ direction, $B_z=B$,
one finds
\beq
\langle {\overline \psi}(z) \, \gamma^0 \, \psi(z) \rangle = -{\delta w
\over \delta A_0} =  {B \over 4 \pi^2} \, {d \over dz} \arctan \left (
{\mu_2(z) \over \mu_1(z)}  \right ) ~.
\label{jslow}
\eeq
The total charge is then given by the total flux of the magnetic field
through the wall, $F = \int dx \, dy \, B_z$ and  the difference of the
phases of the mass term $\mu_1 + i \mu_2$ at two infinities in $z$,
$\Delta \Phi = \arctan (\mu_2/\mu_1)|_{z \to +\infty}-\arctan
(\mu_2/\mu_1)|_{z \to -\infty}$,
\beq
Q=\int   dx \, dy \, dz \,\langle {\overline \psi}(z) \, \gamma^0 \,
\psi(z) \rangle =  {F\, \Delta \Phi \over 4 \pi^2} .
\label{totq}
\eeq

The induced charge is a direct analog of the (generally fractional)
fermionic charge of a kink in (1+1) dimensional models \cite{jr,gw}.
This phenomenon in (3+1) dimensions was considered \cite{ps} in
connection with a magnetic monopole (dyon) traversing an axionic domain
wall, in which process the net change (reversal) of the magnetic flux
across the wall results in charge exchange between the dyon and the
wall\footnote{The paper \cite{ps} contains a reference to an unpublished
communication with H. Georgi and J. Polchinski, who apparently had also
interpreted the effect of the triangle graph  in terms of an induced
charge in a magnetic field. However, as they both kindly corresponded to
me, they had never pursued this issue beyond an unwritten remark. The
monopole -- axion wall charge exchange was  also later discussed by I.
Kogan \cite{ik}.}.

In the specific model, considered in the present paper, the calculation
of the induced charge density automatically comes along with the
calculation of the (eventually vanishing) spontaneous magnetization. It
will be shown that, as expected on general grounds, the relation
(\ref{totq}) between the total induced charge and the total magnetic
flux does not depend on the specific shape of the dependence of the
phase $\Phi$ on $z$. However the {\it distribution} of the induced
charge density in $z$ does depend on the specific rate of variation of
the mass parameter $\mu(z)$, and generally differs from that given by
the equation (\ref{jslow}) that is justified in the limit where the rate
of variation can be considered as slow. One can notice in this
connection that the charge distribution may be of a greater physical
relevance than the total charge, since barring the existence of
monopoles it is physically impossible to produce a net magnetic flux
through an infinite or closed wall. Therefore the total charge has to be
zero, while the distribution of the density of charge can be nontrivial.

The further material in this paper is organized as follows. In Sec.2 the
considered class of models is described as well as some properties of
the relevant operators corresponding to the motion in the $x-y$ plane in
magnetic field and to motion in the $z$ direction in the domain wall
background are discussed. In Sec.3 the dependence on $B$ of the energy
of the ground state of the fermion field is calculated and the vanishing
of all odd terms in the expansion in $B$ is demonstrated including the
vanishing of the spontaneous magnetization. In  Sec.4 a general
expression for the induced charge density is presented, and Sec.5
contains the calculation of this density in the limit of slowly varying
phase as well as a discussion of the topological nature of the total
induced charge. An explicit calculation of the distribution of the
induced charge in two particular sample situations is considered in
Sec.6. Finally, a general discussion and a summary of results are
presented in Sec. 7.

\section{The model and the relevant operators}
The simplifying assumption in the class of models to be considered here
is that the real part $\mu_1$ of the mass term is fixed and nonzero,
while the imaginary part $\mu_2$ depends on $z$. To the best of the
author's knowledge, a model of such type with $\mu_2$ being proportional
to a spinless field $\phi$, $\mu_2=g \, \phi$, with the field varying as
$\phi(z)= v \, \tanh (m_\phi \, z/2)$ across the domain wall, was first
suggested by T.D. Lee and G.C. Wick \cite{lw} as a model of spontaneous
breaking of the $CP$ symmetry. The fermion spectrum and scattering
states in the presence of the domain wall in this model were studied in
Ref. \cite{mv2}.

In a generic case of a non-zero change of $\mu_2$ between the infinities
in $z$ and a non-zero constant $\mu_1$, one can assume for definiteness
without further loss of generality that $\mu_1$ is positive, and
$\mu_2(z)$ changes from a negative value at $z \to -\infty$ to a
positive one at $z \to +\infty$. Also for definiteness it is assumed
here
that a uniform positive magnetic field $B$ is applied in the $z$
direction. In what follows the gauge for the electromagnetic field is
fixed in a standard way such that the vector potential for the field $B$
is given by $A_x=0$, $A_y=B \, x$. Adopting also the standard
representation for the $\gamma$ matrices, the one-particle Hamiltonian,
corresponding to the Lagrangian (\ref{lgen}) takes the following form
\beq
H=\left ( \begin{array}{cccc}
\mu_1 & 0 & i \, P^\dagger & -i \, R \\
0 & \mu_1 &  i \, R^\dagger & i \, P \\
-i \, P & -i \, R & -\mu_1 & 0 \\
i \, R^\dagger & -i \, P^\dagger & 0 & -\mu_1
\end{array}
\right )~,
\label{ham}
\eeq
and $H^2$ has the diagonal form:
\beq
H^2= {\rm diag}\left(\mu_1^2+P^\dagger P+ R R^\dagger, \, \mu_1^2+P
P^\dagger+ R^\dagger R, \, \mu_1^2+P P^\dagger+ R R^\dagger,
\,\mu_1^2+P^\dagger P+ R^\dagger R \right )~.
\label{ham2}
\eeq
These formulas make use of the following notation for the operators
describing respectively the Landau quantization of the motion in the
$x-y$ plane and the quantization of motion along the $z$ axis:
\beq
R=\partial_x + B \, x + p_y,~~ R^\dagger=-\partial_x + B \, x + p_y~,
\label{defr}
\eeq
and
\beq
P=\partial_z + \mu_2(z),~~ P^\dagger=-\partial_z + \mu_2(z)~.
\label{defp}
\eeq
The quantity $p_y$ is the value of the conserved momentum in the $y$
direction (as a consequence of the chosen gauge condition). The energy
levels, determined by the eigenvalues of $R^\dagger R$ and $R R^\dagger$
do not depend on $p_y$ and the degeneracy number is well known to be
given by $B \, S/(2 \pi)$, where $S$ is the normalization area in the
$x-y$ plane. In what follows this degeneracy factor will be explicitly
accounted for, and the value of $p_y$ set to $p_y=0$ in the definition
of the operators (\ref{defr}).

According to Eq.(\ref{ham2}) the spectrum of one-particle energies is
determined by eigenvalues of the operators $R^\dagger R$, $R R^\dagger$,
$P^\dagger P$, and $P P^\dagger$. These spectra exhibit two separate
structures found in supersymmetric quantum mechanics, well known for the
operators $R$ and $R^\dagger$ and also recently used for the $P$ type
operators in connection with kinks in (1+1) dimensional models
\cite{svv}. Namely, if the issue of boundary conditions in $z$ is
ignored, one would naively conclude that the spectra of eigenvalues of
$P^\dagger P$ and $P P^\dagger$ coincide except for an extra zero
eigenvalue (under the adopted sign conventions) of $P^\dagger P$.
Indeed, let $v_k$, $k=1,\,2, \ldots$, be the normalized eigenfunction
corresponding to the (necessarily positive) eigenvalue $\lambda_k^2$ of
the positive operator $P P^\dagger$, so that
\beq
P P^\dagger \, v_k =\lambda_k^2 \, v_k~.
\label{eigv}
\eeq
Applying the operator $P^\dagger$ to both sides of this equation, one
finds that the function
\beq
u_k = P^\dagger \, v_k /\lambda_k
\label{uv}
\eeq
is the normalized eigenfunction of the operator $P^\dagger P$ with the
same eigenvalue $\lambda_k^2$. Applying the operator $P$ to both sides
of the latter relation and using (\ref{eigv}), one find the inverse of
the relation (\ref{uv})
\beq
v_k = P \, u_k /\lambda_k~.
\label{vu}
\eeq
This construction does not work however for the zero mode $u_0$ of
$P^\dagger P$, satisfying the equation $P \, u_0=0$ (BPS state). The
explicit form of the normalizable function $u_0(z)$ is readily found
from the definition (\ref{defp}):
\beq
u_0(z)= {\rm const} \cdot \exp \left (-\int^z \, \mu_2({\tilde z}) \,
d {\tilde z} \right )~.
\label{u0}
\eeq

The operators $R^\dagger$ and $R$ coincide, up to normalization, with
the creation and annihilation operators for a harmonic oscillator, and
their spectra are the text-book ones: the spectrum of eigenvalues of $R
R^\dagger$ is given by $2\, B \, n$ with $n=1,\,2, \ldots$, while that
of $R^\dagger R$ is given by the same simple expression, however also
including $n=0$.

This discussion of the properties of the operators involved in the
Hamiltonian in Eq.(\ref{ham}) is helpful in considering the spectrum of
the one-particle energies. In particular, one can separately consider
each eigenmode of the motion in the $z$ direction as a (2+1) dimensional
fermion system. Then the eigenmodes of the operator $P^\dagger P$
correspond to such system with positive mass parameter $m$:
$m=\sqrt{P^\dagger P + \mu_1^2}$, while the eigenmodes of $P P^\dagger$
correspond to negative $m$: $m=-\sqrt{P P^\dagger + \mu_1^2}$. For the
former modes the negative energy spectrum of Landau levels is given by
$-\sqrt{P^\dagger P + \mu_1^2+ 2 \, B \, n}$ and includes $n=0$, while
for the latter ones the negative energies are given by $-\sqrt{P
P^\dagger + \mu_1^2+ 2 \, B \, n}$ excluding $n=0$. Since all the
non-zero eigenvalues of $P^\dagger P$ and $P P^\dagger$ coincide, one
might very naively conjecture that their effects in magnetization cancel
and the net result is given by only one `unpaired' zero mode of the
operator $P^\dagger P$. Such conjecture however would be false, since
the summation over the modes is generally divergent, and one should
perform a proper calculation with a proper regularization. Also the
discretization of the continuum spectra of the operators $P^\dagger P$
and $P P^\dagger$ requires imposing conditions at the boundaries of
large but finite bounding box in $z$. These conditions generally are not
satisfied by the relations (\ref{uv}) and (\ref{vu}), and there arises a
splitting of the spectra of continuum modes of the operators $P^\dagger
P$ and $P P^\dagger$ which should be accounted for in a proper
calculation.

\section{Dependence of the fermion vacuum energy on $B$.}
For a full regularized calculation of the energy of the ground state of
the fermion field we use here the standard four dimensional technique
with the Pauli-Villars regularization procedure. The latter
regularization preserves the gauge invariance and, importantly for the
discussed problem, the Lorentz covariance\footnote{The vanishing of
spontaneous magnetization can in fact be viewed as due to the
possibility of using a regularization preserving the Lorentz symmetry.}.
This amounts to introducing the regulator fermion field $\Psi$ with
large but finite mass, so that the mass term for $\Psi$ can be written
as $M+i \mu_2 \, \gamma^5$ instead of $\mu_1 + i \mu_2 \, \gamma^5$ for
the `physical' fermion $\psi$, and treat the loop with the regulator
field with an extra minus sign\footnote{Only one regulator field is
indicated here for simplicity of expressions, whereas the full
regularization of the vacuum energy requires additional regulator
fields.
However for the considered here effects in the energy one regulator is
in fact sufficient.}. The regularized expression for the total energy
$W$ then reads as
\begin{eqnarray}
&&W=i \int_{-\infty}^\infty {d p_0 \over 2 \pi} \left \{  {\rm Tr} \ln
\left  [ i D_\alpha \gamma^\alpha - \mu_1 - i \mu_2 (z) \, \gamma^5
\right ] - {\rm Tr} \ln \left  [ i D_\alpha \gamma^\alpha - M - i \mu_2
(z) \, \gamma^5 \right ] \right\} = \nonumber \\
&&i \int_{-\infty}^\infty {d p_0 \over 2 \pi} {\rm Tr} \int_{\mu_1}^M d
m \, \left [ i D_\alpha \gamma^\alpha - m - i \mu_2 (z) \, \gamma^5
\right ]^{-1} ~,
\label{trgen}
\end{eqnarray}
where $i D_\alpha \gamma^\alpha=  \gamma^0 p_0 - \mbox{\boldmath
$\gamma$}\cdot ({\bf p}+{\bf A})$ with the spatial momentum ${\bf p}$
understood as operator, and the trace running over the spinor
indices and the spatial variables.

The inverse of the Dirac operator in the last expression in
Eq.(\ref{trgen}) can be readily found by the usual multiplication of the
numerator and the denominator by $i D_\alpha \gamma^\alpha + m - i \mu_2
(z) \, \gamma^5$. After taking into account the previously mentioned
degeneracy in the momentum $p_y$, the result can be written as
\beq
 W=i {B \, S \over 2 \pi} \int_{-\infty}^\infty {d p_0 \over 2 \pi} {\rm
Tr} \int_{\mu_1}^M d m  \left ( \begin{array}{cccc}
p_0+ m & 0 & -i \, P^\dagger & i \, R \\
0 & p_0+m  &  -i \, R^\dagger & -i \, P \\
-i \, P & -i \, R & -p_0+m & 0 \\
i \, R^\dagger & -i \, P^\dagger & 0 & -p_0+m
\end{array}
\right )  \left [ p_0^2 - H^2(m) \right ]^{-1},
\label{trexpl}
\eeq
where $H^2(m)$ is the same diagonal matrix as in Eq.(\ref{ham2}) with
$\mu_1$ being replaced by $m$. The trace over the spinor variables
leaves only the contribution of the diagonal factors in
Eq.(\ref{trexpl}), which thus can be rewritten as
\beq
W=i {B \, S \over 2 \pi} \int_{-\infty}^\infty {d p_0 \over 2 \pi}
\int_{\mu_1}^M d m \, \left \{{\rm Tr} \left [  {p_0 \gamma^0 \over
p_0^2 - H^2(m)}  \right ] +  {\rm Tr} \left [ {m \over p_0^2 - H^2(m)}
\right ] \right \}~.
\label{trexplm}
\eeq
The first term in the curly braces is manifestly odd in $p_0$, thus the
integration over $p_0$ from $-\infty$ to $\infty$ yields zero. (The
integral can be verified to be finite, thus no further regularization,
that potentially could invalidate the symmetry argument,  is needed.)
The second term after explicitly performing the trace over the spinor
variables takes the form
\begin{eqnarray}
W\!\! \!&=&\!\! \!-i {B \, S \over 2 \pi} \int_{-\infty}^\infty {d p_0
\over 2 \pi}  \int_{\mu_1}^M d m \, m \, {\rm Tr}\left [ \left ({1 \over
-p_0^2+ m^2 +P^\dagger P + R R^\dagger }+ {1 \over -p_0^2+m^2 +
P^\dagger P + R^\dagger R} \right )  \right. \nonumber \\
 & + &\! \! \!\left.\left ({1 \over -p_0^2+m^2 +P P^\dagger + R
R^\dagger }+ {1 \over -p_0^2+m^2 + P P^\dagger + R^\dagger R} \right )
\right ]~.
\label{mterm}
\end{eqnarray}
Here in braces are grouped together the terms with the same ordering of
$P$ and $P^\dagger$ and different order of $R$ and $R^\dagger$. Each of
the two expressions in the braces has the general form
\beq
{\rm Tr} \left ( {1 \over X+R R^\dagger}+ {1 \over X + R^\dagger R}
\right )~,
\label{agen}
\eeq
where the operator $X$ does not depend on the magnetic field. It is now
easy to show that the expression (\ref{agen}) contains only odd powers
in its expansion in $B$ and thus that the expansion in $B$ of the vacuum
energy described by Eq.(\ref{mterm}) contains only even powers, as
required by the $C$ invariance. Indeed, applying the formula
$x^{-1}=\int_0^\infty \exp (-\beta \, x) \, d\beta$ and using the
described spectra of the operators $R R^\dagger$ and $R^\dagger R$, one
can perform the trace over the space of the latter operators and find
the expression (\ref{agen}) in the form
\beq
\int_0^\infty \, d\beta \, \coth \left ( \beta B  \right )\, {\rm Tr}
\exp \left( -\beta X \right )~,
\label{bodd}
\eeq
which is manifestly odd in $B$\,\footnote{Clearly the first term of the
expansion of the expression in Eq.(\ref{bodd}), proportional to
$B^{-1}$, results in a divergent expression, which corresponds to the
divergence of the vacuum energy at $B=0$. This is the only place where
additional
Pauli-Villars regulators are formally required. The subsequent
terms of the expansion however give a finite difference $W(B)-W(0)$ that
is even in $B$.}.

This concludes the proof by an explicit calculation of the absence of
odd powers of the magnetic field $B$ in the expansion of the energy of
the fermion vacuum.

\section{Fermion charge density induced by magnetic field}

We now proceed to a calculation of the electric charge induced by the
external magnetic field $B$. The general formula for the charge density
$\rho(z)=\langle {\overline \psi}(z) \, \gamma^0 \, \psi(z) \rangle$
is obtained from the variational derivative of the energy in
Eq.(\ref{trgen}) with respect to $A_0(z)$ at $A_0=0$ and
reads as
\beq
\rho(z)=-{i \over S} \int_{-\infty}^\infty {d p_0 \over 2 \pi} \left
\langle z \left | {\rm Tr}\left\{ \gamma_0  \,  \left [ i D_\alpha
\gamma^\alpha - \mu_1 - i \mu_2 (z) \, \gamma^5 \right ]^{-1} \right\}
\right |z \right \rangle~.
\label{rhogen}
\eeq
The charge density is finite thus eliminating the need for considering
the contribution of the Pauli-Villars regulator. The trace here runs
over the spinor indices and over the space of the operators $R$ and
$R^\dagger$. The $z$ dependence is thus left `untraced', as is indicated
in Eq.(\ref{rhogen}) by the diagonal matrix element of the remaining
(after `tracing out' the spinor variables and the motion in the $x-y$
plane) $z$ dependent operator in the coordinate representation.
Performing also the trace over the $z$ dependence, i.e. calculating the
integral over $z$, would yield the total induced charge $Q$. Using the
inverse of the Dirac operator as in Eq.(\ref{trexpl}) one can calculate
the trace over the spinor indices and rewrite the latter formula for
$\rho$ as
\begin{eqnarray}
\rho(z)\!\! &=&\!\!  \mu_1 {B  \over 2 \pi} \int_{-\infty}^\infty {d p_0
\over 2 \pi} \left\langle z \left | {\rm Tr}\left [ \left ({1 \over
p_0^2+ \mu_1^2 +P^\dagger P + R^\dagger R }- {1 \over p_0^2+\mu_1^2 +
P^\dagger P + R R^\dagger} \right )  \right. \right. \right.\nonumber \\
 & - &\! \! \left. \left. \left.\left ({1 \over p_0^2+\mu_1^2 +P
P^\dagger + R^\dagger R }- {1 \over p_0^2+\mu_1^2 + P P^\dagger + R
R^\dagger} \right ) \right ] \right | z \right \rangle~.
\label{rho1}
\end{eqnarray}
In the latter expression also the Wick rotation $p_0 \to i p_0$ is done
and the part of the integrand that is odd in $p_0$ is discarded. The
trace over the space of the operators $R$ and $R^\dagger$ in
Eq.(\ref{rho1}) is finite and trivial:
\beq
{\rm Tr}_{(R)} \left ({1 \over X+R^\dagger R}- {1 \over X+R R^\dagger}
\right )={1 \over X}~.
\label{triv}
\eeq
Using this formula in Eq.(\ref{rho1}) one finds
\beq
\rho(z)=\mu_1 {B  \over 2 \pi} \int_{-\infty}^\infty {d p_0 \over 2 \pi}
\left\langle z \left | {1 \over p_0^2+ \mu_1^2 +P^\dagger P }- {1 \over
p_0^2+\mu_1^2 + P P^\dagger} \right | z \right \rangle ~.
\label{rho2}
\eeq
One can notice the absence of higher than linear terms in the expansion
of the charge density in powers of $B$, which is a consequence of the
relation (\ref{triv}). It can be also observed that up to the factor $B
/(2\pi)$ the expression (\ref{rho2}) coincides with similarly calculated
fermionic charge in a (1+1) dimensional theory with $z$ dependent mass
term $\mu_1+i \, \mu_2(z) \, \gamma^5$. This relates the present
calculations to the problem of fermion charge of kinks \cite{jr,gw}.

\section{Charge density for slowly varying phase and the topological
nature of the total charge}

We now consider few specific cases of the dependence of $\mu_2$ on $z$,
in which the charge density can be calculated from Eq.(\ref{rho2})
either in a closed form, or at least `in quadratures'. Also the
topological nature of the total charge is to be addressed in this
section.

The first case to be considered is the one where the rate of variation
of the parameter $\mu_2$ can be considered as slow. In this case in
order to calculate the charge density at a point $z_0$, one can
approximate  $\mu_2(z)$ near $z_0$ by the linear dependence: $\mu_2(z)
\approx \mu_2(z_0)+ (z-z_0) \, \mu_2^\prime(z_0)$ and also expand the
difference of the Green's functions in Eq.(\ref{rho2}) to the linear
order in
$\mu_2^\prime(z_0)$:
\begin{eqnarray}
\rho(z_0)&=&2 \,\mu_1 \, \mu_2^\prime(z_0)\, {B \over 2 \pi}
\int_{-\infty}^\infty {d p_0 \over 2 \pi} \left\langle z_0 \left | \left
(-\partial_z^2 +p_0^2+ \mu_1^2+\mu_2^2(z_0) \right )^{-2} \right | z_0
\right \rangle \nonumber \\
&=&2 \,\mu_1 \, \mu_2^\prime(z_0)\, {B  \over 2 \pi}
\int_{-\infty}^\infty {d p_0 \, d p_z \over (2 \pi)^2} \, \left(
p_z^2+p_0^2 + \mu_1^2+\mu_2^2(z_0) \right )^{-2} \nonumber \\
&=& {B \over 4 \pi^2} \, {\mu_1 \, \mu_2^\prime(z_0) \over
\mu_1^2+\mu_2^2(z_0)}={B \over 4 \pi^2} \, \left. {d \over dz} \arctan
\left( {\mu_2(z) \over \mu_1 } \right ) \right |_{z=z_0} ~,
\label{rslow}
\end{eqnarray}
where the final expression matches that in Eq.(\ref{jslow}) in the
considered case of a constant $\mu_1$. Clearly, for this calculation to
be justified the characteristic length of variation of $\mu_2(z)$ should
be much longer than the `local' Compton wavelength
$(\mu_1^2+\mu_2^2(z_0))^{-1/2}$.

Let us address now the statement that the total integral over the
density,  $Q=\int_{-\infty}^\infty \, \rho(z) \, dz$, is in fact
topological, i.e. it is determined only by the limiting values of
$\mu_2(z)$ at the infinities and does not depend on the specific shape
of the function $\mu(z)$ at finite $z$. From Eq.(\ref{rho2}) the total
integral can be written as a trace:
\begin{eqnarray}
Q &=& \mu_1 {B \, S \over 2 \pi} \int_{-\infty}^\infty {d p_0 \over 2
\pi}\, {\rm Tr} \left ({1 \over p_0^2+ \mu_1^2 +P^\dagger P }- {1 \over
p_0^2+\mu_1^2 + P P^\dagger} \right ) \nonumber \\
&=&  \mu_1 {B \, S \over 4 \pi} {\rm Tr} \left ( {1 \over \sqrt{ \mu_1^2
+P^\dagger P }}- {1 \over \sqrt{\mu_1^2 + P P^\dagger}} \right )~.
\label{trq}
\end{eqnarray}

In the latter expression the traces of the individual operator terms in
the braces are infinite. For this reason one cannot calculate them
separately, each in its own basis. If each trace were finite, one
could calculate the trace of the first operator in the basis of the
eigenfunctions $u_k$ of the operator $P^\dagger P$ and the trace of the
second operator in the basis of the eigenfunctions $v_k$ of the operator
$P P^\dagger$ (cf. equations (\ref{eigv} - \ref{vu})). In that case the
difference would arise only from the `extra' zero mode (\ref{u0}) of the
operator $P^\dagger P$. In view of the divergence of each of the traces
only the trace of the whole operator in braces in Eq.(\ref{trq}) can be
calculated in an arbitrary complete basis. Also the continuum spectrum
of the operators should be discretized by choosing a large bounding box
in $z$: conventionally from $-L/2$ to $+L/2$, where $L$ is sufficiently
large for the asymptotic values of $\mu_2(z)$ to set on, and by imposing
boundary conditions at $z=\pm L/2$. Consider, for instance, the case
where the trace in Eq.(\ref{trq}) is calculated in the basis of the
eigenfunctions $v_k$ of the operator $P P^\dagger$ with an
(anti)periodic condition at the boundaries of the box. Then the
eigenfunctions $u_k$ of the operator $P^\dagger P$ found from
Eq.(\ref{uv}) do not satisfy the same condition since $\mu_2$ takes
different values at the infinities. This effect thus gives rise to a
splitting of the eigenvalues in the two terms in the braces in
Eq.(\ref{trq}). The latter effect however vanishes, if $\mu_2(z)$ takes
the same value at both infinities. Also in this case the operator
$P^\dagger P$ no longer has the zero mode described by Eq.(\ref{u0}),
and the spectra of the two discussed terms in Eq.(\ref{trq}) coincide in
the same basis of (anti)periodic eigenfunctions. Thus the total charge
vanishes, if there is no net variation of $\mu_2$:
$\mu_2(-\infty)=\mu_2(\infty)$.

The latter observation in combination with Eq.(\ref{rslow}) is in fact
sufficient to conclude that the total charge depends only on the
difference of the limiting values of $\mu_2$. Indeed, consider a large
box, defined by the length $L$, and  let the difference $\Delta
=\mu_2(L/2)-\mu(-L/2)$ be nonzero. Consider now an {\it extended}
system, where the box is continued from $z=L/2$ to $z={\tilde L}$, in
such a way that the length of the extension ${\tilde L}-L/2$ is
arbitrarily large, and the behavior of $\mu_2(z)$ in the extension is
chosen `by hand' such that it is a slow function smoothly interpolating
$\mu_2$ from its fixed value at $z=L/2$ back to $\mu_2(-L/2)$ at
$z={\tilde L}$. In other words, the net variation over the extension
${\tilde \Delta} = \mu_2({\tilde L})-\mu_2(L/2)$ is  ${\tilde
\Delta}=-\Delta$. The net change of $\mu_2$ over all the extended box is
then zero. Thus the total charge of the extended system is also zero.
The total charge however is a sum of the one inside the original box,
$Q$, and that in the extension, ${\tilde Q}$. Thus $Q=-{\tilde Q}$. On
the other hand, the charge ${\tilde Q}$ can be calculated in the limit
of slowly varying $\mu_2(z)$ from the formula (\ref{rslow}). Therefore
one arrives at the equation (\ref{totq}) for the charge $Q$,
independently of the details of the behavior of $\mu_2(z)$ at
intermediate $z$ within the original bounding box.

\section{Charge density in sample models}
Here we consider explicit calculations of $\rho(z)$ from the formula
(\ref{rho2}) in sample models of the dependence of $\mu_2(z)$ on $z$ in
order to illustrate that the distribution of charge generally does not
follow the slow variation limit described by Eq.(\ref{rslow}) while the
total charge is of course given by the universal formula (\ref{totq}).

A limit that is maximally opposite to that of a slow variation is where
$\mu_2$ changes as a step function at a point in $z$. Here we consider
the situation where $\mu_2$ changes at $z=0$ from a negative constant
value $\mu_2(z)=-\mu_2$ at $z < 0$ to a positive one, $\mu_2(z)=\mu_2$
at $z > 0$. The Green's functions in Eq.(\ref{rho2}) can be found
explicitly by the standard method of matching at $z=0$:
\begin{eqnarray}
\left\langle z \left | {1 \over p_0^2+ \mu_1^2 +P^\dagger P } \right | z
\right \rangle &=&{1 \over 2 \, q}\, \left [1 + {\mu_2 \over q - \mu_2}
\, \exp \left ( - 2 \, q \,|z| \right ) \right ]~, \nonumber \\
\left\langle z \left | {1 \over p_0^2+ \mu_1^2 +P P^\dagger } \right | z
\right \rangle &=&{1 \over 2 \, q}\, \left [1 - {\mu_2 \over q + \mu_2}
\, \exp \left ( - 2 \, q \,|z| \right ) \right ]~,
\label{resolvs}
\end{eqnarray}
where $q=\sqrt{p_0^2+\mu_1^2+\mu_2^2}$. The charge density is then given
`in quadratures' by the formula
\beq
\rho(z)=\mu_1 \, \mu_2 \, {B  \over 2 \pi} \int_{-\infty}^\infty {d p_0
\over 2 \pi}\, { \exp \left ( - 2 \, \sqrt{p_0^2+\mu_1^2+\mu_2^2} \,|z|
\right ) \over \left ( p_0^2 + \mu_1^2 \right ) }~,
\label{rstep}
\eeq
which, as one could expect, describes an exponential distribution around
the discontinuity point $z=0$ over the range $(\mu_1^2+\mu_2^2)^{-1/2}$
that is the Compton wavelength of the fermion. For the total charge the
integration over $z$ and then over $p_0$ is elementary, and the result
matches the general formula (\ref{totq}).

It can be also noted, in connection with this example, that the
non-exponential term in individual Green's functions in
Eq.(\ref{resolvs}) would lead to a divergence upon integration over
$p_0$. However this term cancels in the difference, leaving the
expression for $\rho(z)$ finite. This illustrates the behavior discussed
in general terms in the previous section.

Another class of specific situations where the Green's functions are
readily calculable in a closed form and the charge density can be found
explicitly is presented by the Lee-Wick model \cite{lw}. In this model
the term $\mu_2$ in the fermion mass arises from the interaction with a
neutral field $\phi$: $\mu_2 = g \, \phi$, with $g$ being the Yukawa
coupling. The $\phi^4$ self-interaction of the field $\phi$ is assumed
to lead to the domain wall solution
\beq
\phi(z)=v \, \tanh \left ( \lambda \, z / 2 \right )~,
\label{twall}
\eeq
which interpolates between the two vacuum states (domains) with
$\phi=\pm v$, and where $\lambda$ is the constant of the
self-interaction, such that the mass of the $\phi$ bosons in either of
the vacua is $m_\phi=\lambda \, v$.

The behavior of the fermion states in such background depends on the
ratio $g/\lambda$ \cite{mv2}. Clearly, the case $g \gg \lambda$
corresponds to the limit of slow variation of $\mu_2$, while  the
opposite case, $g \ll \lambda$, corresponds to the approximation of an
abrupt change in $\mu_2$, provided that in the latter case $\mu_1$ is
also assumed to be small in comparison with $m_\phi$. For a generic
value of the ratio $g/\lambda$ the operators $P^\dagger P$ and $P
P^\dagger$ correspond to the solvable potentials of the form ${\rm
const}/\cosh^2(\lambda \, z / 2)$, and the Green's functions can be
found in terms of hypergeometric functions. In a situation where $2 \,
g/\lambda$ is an integer: $2 \, g/\lambda=N$, both potentials are non
reflecting, and the algebra is greatly simplified, since the relevant
hypergeometric functions collapse to  (Jacobi) polynomials in $\tanh
(\lambda \, z /2)$ of the power $N$ for $P^\dagger P$ and $N-1$ for $P
P^\dagger$. It can be also noted, that the case of $N=2$ in the
considered class of models would correspond to a supersymmetric model,
albeit with $\mu_1=0$, which would take us beyond the assumptions
adopted in the present paper. Here for illustrative purposes we consider
only the most algebraically simple case of $N=1$.

For $\lambda=2 \, g$, the operator $P P^\dagger$ corresponds to a
constant potential: $P P^\dagger=-\partial_z^2 + g^2 \, v^2$, and the
corresponding Green's function is especially simple:
\beq
G_{(P P^\dagger)}(z_1, z_2;\, p_0^2+\mu_1^2) \equiv \left\langle z_1
\left | {1 \over p_0^2+\mu_1^2+P P^\dagger } \right | z_2 \right \rangle
= { \exp \left(-q \, |z_1-z_2| \right) \over 2 \, q}~,
\label{gp}
\eeq
with $q=\sqrt{p_0^2+\mu_1^2+g^2 \, v^2}$. The Green's function for the
operator
$P^\dagger P$ can then be found \cite{svv}, using the relations
(\ref{uv}) and (\ref{u0}) for its eigenfunctions\footnote{The individual
Green's functions are finite, thus each can be expanded in the basis of
the corresponding eigenfunctions.}:
\begin{eqnarray}
&&G_{(P^\dagger P)}(z_1, z_2;\, p_0^2+\mu_1^2) \equiv \left\langle z_1
\left | {1 \over p_0^2+\mu_1^2+P^\dagger P } \right | z_2 \right \rangle
= \sum_{k=0}^\infty \, {u_k(z_1) \, u_k^*(z_2) \over p_0^2+\mu_1^2+
\lambda_k^2} \nonumber \\
&&=\sum_{k=1}^\infty \, P_{z_1}^\dagger \, {v_k(z_1) \, v_k^*(z_2)
\over \lambda_k^2 \, (p_0^2+\mu_1^2+ \lambda_k^2) } \, P_{z_2}+
{u_0(z_1) \, u_0(z_2) \over p_0^2+\mu_1^2}
\label{gm} \\
&&={1 \over p_0^2+\mu_1^2} \left [ P^\dagger \, G_{(P P^\dagger)}(z_1,
z_2;\, 0)\, P -P^\dagger \, G_{(P P^\dagger)}(z_1, z_2;\,
p_0^2+\mu_1^2)\, P \right] + {u_0(z_1) \, u_0(z_2) \over
p_0^2+\mu_1^2}~. \nonumber
\end{eqnarray}
Applying in this formula the operators $P^\dagger$ and $P$ as indicated
to the expression (\ref{gp}), and using the zero mode from
Eq.(\ref{u0}): $u_0(z)=\sqrt{gv/2}/\cosh (gvz)$, one finds that at
coinciding points the Green's function (\ref{gm}) is given by
\beq
G_{(P^\dagger P)}(z, z;\, p_0^2+\mu_1^2)={1 \over 2 \, q} + {g^2 \, v^2
\over 2 \, q \, (p_0^2+\mu_1^2) \, \cosh^2 (g\,v\,z) }~,
\label{gmm}
\eeq
where, again, $q=\sqrt{p_0^2+\mu_1^2+g^2 \, v^2}$.

Upon substitution of the results from the equations (\ref{gp}) and
(\ref{gmm}) to Eq.(\ref{rho2}) for the charge density, and after the
integration over $p_0$, the final expression for the distribution of
charge in this model is found in a remarkably simple explicit form:
\beq
\rho(z)=  {g \, v \over 2 \, \cosh^2 (g\,v\,z)} \,{Q \over S} ~,
\label{rtr}
\eeq
where $Q=B \, S \arctan [gv/(\mu_1^2 + g^2 v^2)]/(2 \pi^2)$ is the total
induced charge, in agreement with the general formula (\ref{totq}).

\section{Discussion}

The explicit calculation described in the sections 2 and 3 illustrates
the implementation of the general statement about the absence of
spontaneous magnetization based on the $C$ parity. In the concrete
calculation the expected cancellation occurs due the fact that it is
possible to regularize the theory while manifestly preserving the charge
conjugation symmetry. Alternatively, as already noted, this cancellation
can be viewed as the manifest Lorentz invariance of the regularized
theory, since a nonzero spontaneous magnetization would certainly imply
a breaking of the Lorentz symmetry \cite{hosotani}. In other words,
neither the $C$ symmetry nor the Lorentz one are anomalous. The same
symmetries preclude an appearance of a spontaneous magnetization also in
models where the fermions are assumed to have anomalous magnetic or/and
electric dipole moments, including the cases \cite{fz} where the
fermions are neutron-like, i.e. with zero charge and nonzero dipole
moments.

The really existing effect, allowed by both these symmetries, is the
fermion charge density induced by an external magnetic field applied
across the domain wall, which can be traced to the well-known anomaly in
the axial current. The total induced charge is of a topological nature,
and is determined, according to Eq.(\ref{totq}), by the total magnetic
flux through the wall and the asymptotic values of the fields in the
model far from the wall. The distribution of the induced charge however,
given by the general formula (\ref{rho2}) depends on the details of the
profile of the wall, as discussed in the sections 5 and 6. This effect,
in principle, can be relevant in detailed analyses of phenomena at the
walls in the early universe -- either for flat domain walls, which could
exist during some epoch, or for dynamics near the walls of bubbles
during a first order phase transition.

The arguments, based on the $C$ parity and/or the Lorentz symmetry, do
not directly apply to an asymmetric state with a net overall fermion
charge, e.g. to the early universe with the baryon asymmetry, that is
neither $C$ nor Lorentz symmetric. In this case a spontaneous
magnetization of the wall is generally allowed, and is proportional to
the asymmetry parameter of the considered state \cite{mv}. It should
however be clearly understood that even a magnetized domain wall does
not produce a magnetic field, which is a simple consequence of the
classical Maxwell's equations \cite{mv}. Therefore domain walls could
not be a source of a correlated at cosmological distances magnetic field
in the early universe, although their electromagnetic properties, like
the charge distribution induced by a magnetic field, could generally be
of importance in other aspects of dynamics of the early universe.

\section{Acknowledgements}

I am thankful to A. Larkin, M. Shifman and A. Vainshtein for
illuminating discussions and to H. Georgi and J. Polchinski for
correspondence related to the hitory of the problem of the induced
charge. This work is supported in part by DOE under the grant number
DE-FG02-94ER40823.

\end{document}